\documentclass[12pt]{iopart}

\usepackage{iopams}
\usepackage{graphicx}

\newcommand{\bra}[1]{\langle{#1}|}
\newcommand{\ket}[1]{|{#1}\rangle}

\begin{document}

\title[Dissipation and entanglement dynamics for two interacting qubits...]
{Dissipation and entanglement dynamics for two interacting qubits
coupled to independent reservoirs}

\author{M. Scala}

\address{Departamento de \'Optica, Facultad de F\'isica, Universidad
Complutense de Madrid, 28040, Spain \\ Dipartimento di Scienze
Fisiche ed Astronomiche dell'Universit\`{a} di Palermo, via
Archirafi 36, I-90123 Palermo, Italy}
\ead{matteo.scala@fisica.unipa.it}

\author{R. Migliore}

\address{CNR-INFM, CNISM and Dipartimento di Scienze Fisiche ed
Astronomiche dell'Universit\`{a} di Palermo, via Archirafi 36,
I-90123 Palermo, Italy}\ead{rosanna@fisica.unipa.it}

\author{A. Messina}

\address{MIUR and Dipartimento di Scienze Fisiche ed
Astronomiche dell'Universit\`{a} di Palermo, via Archirafi 36,
I-90123 Palermo, Italy} \ead{messina@fisica.unipa.it}
\begin{abstract}

We derive the master equation of a system of two coupled qubits by
taking into account their interaction with two independent bosonic
baths. Important features of the dynamics are brought to light,
such as the structure of the stationary state at general
temperatures and the behaviour of the entanglement at zero
temperature, showing the phenomena of sudden death and sudden
birth as well as the presence of stationary entanglement for long
times. The model here presented is quite versatile and can be of
interest in the study of  both Josephson junction architectures
and cavity-QED.
\end{abstract}



\maketitle

\section{Introduction}

During the last two decades the problem of controlling the coherent
time evolution of quantum systems has received a great deal of
attention both because it could lead to a deeper understanding of
fundamental aspects of quantum mechanics and because it is of
interest for applications \cite{harochebook}. The possibility of
generating and controlling non classical correlations (i.e.
entangled states) in multipartite systems, despite their coupling
with an external environment, is a fundamental ingredient for
instance in the theory of quantum measurement and in the study of
the border between the quantum world and the macroscopic classical
world \cite{wheelerzurekbook}. In addition such a possibility is an
essential goal in the field of quantum computing and quantum
information theory \cite{nielsenchuang}. It is well known that the
interaction of an open quantum system with an external reservoir is
an important source of dissipation and decoherence. In order to
describe such phenomena, a master equation approach can be used
\cite{cohen_libro,gardiner,petruccionebook}. In particular,
following Ref. \cite{petruccionebook}, one has at his disposal a
general formalism allowing to derive the quantum master equation of
a general quantum system, provided one knows the Hamiltonian
governing the unitary part of its dynamics. Exploiting this recipe,
one can see that the dissipative dynamics, in the Born-Markov and
rotating wave approximations, is described by a master equation in
which the quantum jumps occur among eigenstates of the Hamiltonian
of the open quantum system under scrutiny. This microscopic approach
may be sometimes in contrast with some phenomenological approaches
to quantum dissipative dynamics present in the literature, as one
can see for example in Refs. \cite{Scala,Scala2,Scala3}.

Within this framework and with the aforementioned approach, here we
analyze the dynamical behavior of the entanglement between two
coupled two-state systems each of them interacting with a bosonic
bath. The model is quite versatile and can be exploited, for
instance, in order to describe the dipole-dipole (flux or charge)
interaction of two distant atomic (flux or charge) qubits. In the
case of Josephson flux qubits the coupling term corresponds to a
flux-flux coupling proportional to their mutual inductance
\cite{JJRosanna}. It is worth noting that the same model can be used
to describe richer physical situations, for instance the coupling
between a Josephson junction based qubit in the charge regime with
an impurity located in the substrate of the device \cite{Paladino}.
In this case the qubit can be thought as coupled with external
environmental degrees of freedom describing the auxiliary circuitry
(necessary, for instance, for the qubit control and the redout)
which can be modeled as an infinite bath of harmonic oscillators,
while the substrate impurity may interact with a phononic bath.

Starting from such a model of qubit-qubit interaction, and
including counter-rotating terms in the system Hamiltonian, we
derive a quantum master equation in order to describe the
dissipative dynamics of the two-qubit system, concentrating on
some aspects such as the structure of the stationary state of the
system and the time evolution of the entanglement between the two
qubits. The entanglement dynamics of open bipartite quantum
systems has been studied in previous works, for example in Refs.
\cite{Yu,bellomo,Tanas}, bringing to light important features such
as the complete disentanglement of the system in a finite time
(entanglement sudden death), and, in the case of Ref.
\cite{Tanas}, the sudden reappearance of entanglement (sudden
birth). In the latter case, the presence of stationary
entanglement for very long times has been related to the presence
of a reservoir common to the two subsystems, so that the non-local
correlations between them can be thought of as due to the
mediation of the environment. In this paper we will show that
stationary entanglement can occur also in the case wherein the two
qubits interact with two independent reservoirs, provided the
interaction between the two subsystems contains also
counter-rotating terms, which are usually neglected in the
literature. In such a case, the stationary entanglement occurs
because the counter-rotating terms cause the ground state of the
system to be an entangled state of the bipartite system.

The paper is structured as follows. In section 2 we introduce the
model previously described and we microscopically derive the
Markovian master equation in the weak damping limit assuming that
the two reservoir are independent and with arbitrary temperatures,
$T_1$ and $T_2$. Therefore, in section 4, we analyze the dynamics
of the system in the limit case $T_1=T_2=0$ discussing the
stationary entanglement and the features of its time evolution
considering different initial conditions for the bipartite system.
Finally, conclusive remarks are given in section 5.

\section{The model}
Let us consider two interacting two-level systems and let us call
$\ket{0}_1$ ($\ket{0}_2$) the ground state of the first (second)
system and $\ket{1}_1$ ($\ket{1}_2$) the corresponding excited
state. Let us assume that the two systems are coupled in such a
way that their unitary dynamics is governed by the following
Hamiltonian (in units of $\hbar$):
\begin{eqnarray}\label{Hamiltonianmodel}
 H_S=\omega_1\sigma_+^{(1)}\sigma_-^{(1)}+\omega_2\sigma_+^{(2)}\sigma_-^{(2)}
 +\frac{\lambda}{2}\,\sigma_x^{(1)}\sigma_x^{(2)},
\end{eqnarray}
where $\omega_i$ is the Bohr frequency of the $i$-th two-level
system, $\lambda/2$ is the coupling constant and where we have used
the Pauli operators
$\sigma_+^{(i)}=\left|1\right>{}_i{}_i\left<0\right|$,
$\sigma_-^{(i)}=\left|0\right>{}_i{}_i\left<1\right|$ and
$\sigma_x^{(i)}= \sigma_+^{(i)}+\sigma_-^{(i)}$, with $i=1,2$. In
the case of Josephson flux qubits the coupling term in Eq.
(\ref{Hamiltonianmodel}) corresponds to a flux-flux coupling with
$\lambda/2$ proportional to their mutual inductance
\cite{JJRosanna}.

It is worth noting that in the Hamiltonian also the counter-rotating
terms of the interaction have been included and we will see that
they play a central role in the dynamics of the entanglement between
the two systems.

The model in Eq. (\ref{Hamiltonianmodel}) can be exactly
diagonalized. By exploiting the fact that the Hamiltonian $H_S$ in
the uncoupled basis
$\left\{\ket{00},\ket{11},\ket{10},\ket{01}\right\}$, where for
instance $\ket{00}=\ket{0}_1\otimes\ket{0}_2$, is block diagonal, it
is straightforward to show that the eigenvalues (given for
increasing energies) are:
\begin{eqnarray}\label{eigenvalues}
 E_a=\frac{1}{2}\left(\omega_1+\omega_2\right)-\frac{1}{2}\sqrt{\left(\omega_2+\omega_1\right)^2+\lambda^2}\nonumber\\
 \nonumber\\
 E_b=\frac{1}{2}\left(\omega_1+\omega_2\right)-\frac{1}{2}\sqrt{\left(\omega_2-\omega_1\right)^2+\lambda^2}\nonumber\\
 \nonumber\\
 E_c=\frac{1}{2}\left(\omega_1+\omega_2\right)+\frac{1}{2}\sqrt{\left(\omega_2-\omega_1\right)^2+\lambda^2}\nonumber\\
 \nonumber\\
 E_d=\frac{1}{2}\left(\omega_1+\omega_2\right)+\frac{1}{2}\sqrt{\left(\omega_2+\omega_1\right)^2+\lambda^2},
\end{eqnarray}
while the corresponding eigenstates are:
\begin{eqnarray}\label{eigenstates}
 \ket{a}=\cos\frac{\theta_I}{2}\ket{00}-\sin\frac{\theta_I}{2}\ket{11}\nonumber\\
 \nonumber\\
 \ket{b}=\cos\frac{\theta_{II}}{2}\ket{10}-\sin\frac{\theta_{II}}{2}\ket{01}\nonumber\\
 \nonumber\\
 \ket{c}=\sin\frac{\theta_{II}}{2}\ket{10}+\cos\frac{\theta_{II}}{2}\ket{01}\nonumber\\
 \nonumber\\
 \ket{d}=\sin\frac{\theta_I}{2}\ket{00}+\cos\frac{\theta_I}{2}\ket{11}.
\end{eqnarray}
Here $\omega_2\ge\omega_1$ and the parameters $\theta_I$ and
$\theta_{II}$ satisfy the relations:
\begin{eqnarray}\label{thetaI}
 \sin\theta_I=\frac{\left|\lambda\right|}{\sqrt{\left(\omega_2+\omega_1\right)^2+\lambda^2}},\;&
 \cos\theta_I=\frac{\omega_1+\omega_2}{\sqrt{\left(\omega_2+\omega_1\right)^2+\lambda^2}}\\
 \label{thetaII}\sin\theta_{II}=\frac{\left|\lambda\right|}{\sqrt{\left(\omega_2-\omega_1\right)^2+\lambda^2}},\;&
 \cos\theta_{II}=\frac{\omega_2-\omega_1}{\sqrt{\left(\omega_2-\omega_1\right)^2+\lambda^2}}
\end{eqnarray}

The losses in the system under scrutiny will be taken into account
by considering the coupling between the $i$-th system and its own
reservoir at temperature $T_i$. In the following we will consider
the case of independent bosonic reservoirs, whose temperatures, in
the general case, can take different values. The total Hamiltonian
of the bipartite system and the reservoirs can thus be written as
follows:
\begin{eqnarray}\label{HTotal}
H=H_S+H_E+H_{\mathrm{int}}\nonumber\\
\nonumber\\
H_E=\sum_k\omega_ka_k^\dag a_k+\sum_j\omega_jb_j^\dag b_j\nonumber\\
\nonumber\\
H_{\mathrm{int}}=\sigma_x^{(1)}\otimes\sum_k\epsilon_k\left(a_k+a_k^\dag\right)
+\sigma_x^{(2)}\otimes\sum_jg_j\left(b_j+b_j^\dag\right).
\end{eqnarray}
Here $a_k$ ($a_k^\dag$) is the annihilation (creation) operator of
the $k$-th mode (of frequency $\omega_k$) of the reservoir
interacting with the first subsystem and similarly $b_j$
($b_j^\dag$) is the annihilation (creation) operator of the $j$-th
mode (of frequency $\omega_j$) of the reservoir interacting with the
second subsystem. The parameters $\epsilon_k$ and $g_j$ are the
corresponding coupling constants.

From this model it is possible to microscopically derive the
Markovian master equation describing all the relaxation phenomena in
the dynamics of the bipartite system under study. To this end in the
next section we will exploit the general formalism given in Ref.
\cite{petruccionebook}.

\section{Derivation of the Markovian master equation in the weak damping limit}

From the model in Eq. (\ref{HTotal}) it is possible to
microscopically derive the master equation for the evolution of
the bipartite system, by following the general procedure outlined
in Ref. \cite{petruccionebook},  in the Born-Markov and rotating
wave approximations\footnote{We stress the point that in this
paper we will always call rotating wave approximation the
operation of neglecting rapidly oscillating terms in the
dissipative part of the master equation. In addition, we note that
this should not be confused with the elimination of the
counter-rotating terms in the Hamiltonian of the system which will
always be taken into account.}. The main point of the formalism is
that all the jump processes involve transitions between dressed
states of the open system under study, i.e. the eigenstates of the
Hamiltonian $H_S$ of the system, which in our case is given by Eq.
(\ref{eigenstates}).

It is well known that, in the Schr\"odinger picture, the Markovian
master equation for a generic open quantum system with Hamiltonian
$H_0$ is given by \cite{petruccionebook}:
\begin{eqnarray}\label{MEqGeneral}
 \dot{\rho}(t)&=&-i\left[H_0,\rho(t)\right]\nonumber\\
 \nonumber\\
 &+&\sum_\omega\sum_{\alpha,\beta}\gamma_{\alpha,\beta}(\omega)
 \left(A_\beta(\omega)\rho(t)A_\alpha^\dag(\omega)
 -\frac{1}{2}\left\{A_\alpha^\dag(\omega)A_\beta(\omega),\rho(t)\right\}\right),
\end{eqnarray}
where the symbol $\left\{\cdot,\cdot\right\}$ denotes the
anticommutator between operators. In the derivation of Eq.
(\ref{MEqGeneral}) we have to assume that the interaction
Hamiltonian between the system and the environment is of the form
$H_{I}=\sum_\alpha A_\alpha\otimes B_\alpha$, where
$A_\alpha=A_\alpha^\dag$ acts on the Hilbert space of the open
system under scrutiny, while $B_\alpha=B_\alpha^\dag$ acts on the
Hilbert space of the environment. In particular, in the case of
the model in Eq. (\ref{HTotal}) the sum consists of two terms,
since we have assumed that each two-level system interacts with
its own reservoir. In Eq. (\ref{MEqGeneral}), a renormalization
Hamiltonian has been neglected and the rates
$\gamma_{\alpha,\beta}(\omega)$ are given by the Fourier
transforms of the environment correlation functions, according to:
\begin{eqnarray}\label{gammagen}
 \gamma_{\alpha,\beta}(\omega)=\int_{-\infty}^{+\infty}d\tau\mathrm{e}^{i\omega\tau}
 \left<B_\alpha^\dag(\tau)B_\beta(0)\right>.
\end{eqnarray}
Concerning the jump operators $A_\alpha(\omega)$, their number is
given by the number of different Bohr frequencies relative to $H_0$
and they can be calculated from the relation \cite{petruccionebook}:
\begin{eqnarray}\label{jumpgen}
 A_\alpha(\omega)\equiv\sum_{\epsilon'-\epsilon=\omega}\Pi(\epsilon)A_\alpha\Pi(\epsilon'),
\end{eqnarray}
where $\Pi(\epsilon)$ is the projector on the eigenspace of the open
system relative to the energy $\epsilon$ and the sum is extended to
all the couples of $\epsilon$ and $\epsilon'$ such that
$\epsilon'-\epsilon=\omega$.

For our model we can identify the operators $A_1=\sigma_x^{(1)}$,
$A_2=\sigma_x^{(2)}$,
$B_1=\sum_k\epsilon_k\left(a_k+a_k^\dag\right)$ and
$B_2=\sum_jg_j\left(b_j+b_j^\dag\right)$. By calculating the matrix
elements of $A_1$ and $A_2$ between the eigenstates of $H_S$ given
in Eq. (\ref{eigenstates}), it is possible to see that there are
only two possible values for the Bohr frequencies of the transitions
allowed. The first one is
$$\omega_I=\frac{1}{2}\left(\sqrt{\left(\omega_2+\omega_1\right)^2
+\lambda^2}-\sqrt{\left(\omega_2-\omega_1\right)^2+\lambda^2}\right),$$
for the transitions $\ket{b}\rightarrow\ket{a}$ and
$\ket{d}\rightarrow\ket{c}$, corresponding to the jump operator:
\begin{eqnarray}\label{JI1}
J_{I1}=\bra{a}A_1\ket{b}\ket{a}\bra{b}+\bra{c}A_1\ket{d}\ket{c}\bra{d}
\end{eqnarray}
relative to the coupling of the first qubit with its own reservoir,
and to:
\begin{eqnarray}\label{JI2}
J_{I2}=\bra{a}A_2\ket{b}\ket{a}\bra{b}+\bra{c}A_2\ket{d}\ket{c}\bra{d}
\end{eqnarray}
due to the coupling between the second qubit with the corresponding
reservoir.

Similarly, the second Bohr frequency is:
$$\omega_{II}=\frac{1}{2}\left(\sqrt{\left(\omega_2+\omega_1\right)^2
+\lambda^2}+\sqrt{\left(\omega_2-\omega_1\right)^2+\lambda^2}\right),$$
for the transitions $\ket{c}\rightarrow\ket{a}$ and
$\ket{d}\rightarrow\ket{b}$, corresponding to the jump operator:
\begin{eqnarray}\label{JII1}
 J_{II1}=\bra{a}A_1\ket{c}\ket{a}\bra{c}+\bra{b}A_1\ket{d}\ket{b}\bra{d}
\end{eqnarray}
relative to the coupling of the first qubit with its own reservoir,
and to:
\begin{eqnarray}\label{JII2}
 J_{II2}=\bra{a}A_2\ket{c}\ket{a}\bra{c}+\bra{b}A_2\ket{d}\ket{b}\bra{d}
\end{eqnarray}
due to the coupling between the second qubit with the corresponding
reservoir.


\begin{figure}
 \begin{center}
     \includegraphics[width=0.4\textwidth]{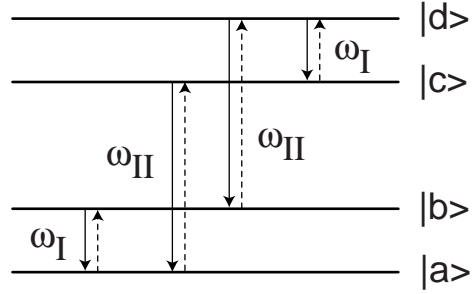}
    \caption{Schematic illustration of the eigenstates of the bipartite qubit-qubit system and
    of the allowed transitions characterized by the Bohr frequencies $\omega_I$ and $\omega_{II}$.}\label{figura1}
 \end{center}
\end{figure}


Inserting in Eq. (\ref{MEqGeneral}) the structure of the jump
operators given in Eqs. (\ref{JI1})-(\ref{JII2}), the Markovian
master equation can be cast in the following form:
\begin{eqnarray}\label{MeqSistema}
 \dot{\rho}(t)&=&-i\left[H_S,\rho(t)\right]\nonumber\\
 \nonumber\\
 &+&\sum_{i=I}^{II}\sum_{l=1}^{2}\gamma_{i,\,ll}\left(J_{il}\rho(t)J_{il}^\dag
 -\frac{1}{2}\left\{J_{il}^\dag J_{il},\rho(t)\right\}\right)\nonumber\\
 \nonumber\\
 &+&\sum_{i=I}^{II}\sum_{l=1}^{2}\bar{\gamma}_{i,\,ll}\left(J_{il}^\dag\rho(t)J_{il}
 -\frac{1}{2}\left\{J_{il} J_{il}^\dag,\rho(t)\right\}\right)
\end{eqnarray}
where, assuming that  the two reservoirs are independent and each of
them is in a thermal state, with temperatures $T_1$ and $T_2$
respectively, one has:
\begin{eqnarray}\label{gammasistema}
 \gamma_{i,\,lm}=\int_{-\infty}^{+\infty}d\tau\mathrm{e}^{i\omega_i\tau}
 \left<B_l^\dag(\tau)B_m(0)\right>,
\end{eqnarray}
and the Kubo-Martin-Schwinger relation
\begin{eqnarray}\label{KMS}
 \bar{\gamma}_{i,\,ll}=\mathrm{e}^{-\omega_i/K_BT_l}{\gamma}_{i,\,ll}
\end{eqnarray}
holds \cite{petruccionebook}. The hypotesis of independent
reservoirs we have made consists in assuming that
$\left<B_l^\dag(\tau)B_m(0)\right>=0$, i.e. $\gamma_{i,\,lm}=0$,
when $l\neq m$.

Finally, by inserting Eqs. (\ref{JI1})-(\ref{JII2}) into Eq.
(\ref{MeqSistema}) and rearranging the terms, we obtain:
\begin{eqnarray}\label{Meqfinale}
 \dot{\rho}(t)=-i\left[H_S,\rho(t)\right]\\
 \nonumber\\
 +c_I\left(\ket{a}\bra{b}\rho(t)\ket{b}\bra{a}-\frac{1}{2}\left\{\ket{b}\bra{b},\rho(t)\right\}\right)
 +c_{II}\left(\ket{a}\bra{c}\rho(t)\ket{c}\bra{a}-\frac{1}{2}\left\{\ket{c}\bra{c},\rho(t)\right\}\right)\nonumber\\
 \nonumber\\
  +c_I\left(\ket{b}\bra{d}\rho(t)\ket{d}\bra{b}-\frac{1}{2}\left\{\ket{d}\bra{d},\rho(t)\right\}\right)
  +c_{II}\left(\ket{c}\bra{d}\rho(t)\ket{d}\bra{c}-\frac{1}{2}\left\{\ket{d}\bra{d},\rho(t)\right\}\right)\nonumber\\
 \nonumber\\
  +\bar{c}_I\left(\ket{b}\bra{a}\rho(t)\ket{a}\bra{b}-\frac{1}{2}\left\{\ket{a}\bra{a},\rho(t)\right\}\right)
  +\bar{c}_{II}\left(\ket{c}\bra{a}\rho(t)\ket{a}\bra{c}-\frac{1}{2}\left\{\ket{a}\bra{a},\rho(t)\right\}\right)\nonumber\\
 \nonumber\\
 +\bar{c}_I\left(\ket{d}\bra{b}\rho(t)\ket{b}\bra{d}-\frac{1}{2}\left\{\ket{b}\bra{b},\rho(t)\right\}\right)
 +\bar{c}_{II}\left(\ket{d}\bra{c}\rho(t)\ket{c}\bra{d}-\frac{1}{2}\left\{\ket{c}\bra{c},\rho(t)\right\}\right)\nonumber\\
 \nonumber\\
 +c_{cr,I}\left(\ket{a}\bra{b}\rho(t)\ket{d}\bra{c}+\ket{c}\bra{d}\rho(t)\ket{b}\bra{a}\right)
 +c_{cr,II}\left(\ket{a}\bra{c}\rho(t)\ket{d}\bra{b}+\ket{b}\bra{d}\rho(t)\ket{c}\bra{a}\right)\nonumber\\
 \nonumber\\
 +\bar{c}_{cr,I}\left(\ket{d}\bra{c}\rho(t)\ket{a}\bra{b}+\ket{b}\bra{a}\rho(t)\ket{c}\bra{d}\right)
 +\bar{c}_{cr,II}\left(\ket{d}\bra{b}\rho(t)\ket{a}\bra{c}+\ket{c}\bra{a}\rho(t)\ket{b}\bra{d}\right).\nonumber
\end{eqnarray}
The decay rates $c_i$, and the cross terms $c_{cr,i}$ (with
$i=I,II$), are given by:
\begin{eqnarray}\label{cI}
 c_I&=&\gamma_{I,11}\left(\cos\frac{\theta_I}{2}\cos\frac{\theta_{II}}{2}
 +\sin\frac{\theta_I}{2}\sin\frac{\theta_{II}}{2}\right)^2\nonumber\\
 \nonumber\\
 &+&\gamma_{I,22}\left(\cos\frac{\theta_I}{2}\sin\frac{\theta_{II}}{2}
 +\sin\frac{\theta_I}{2}\cos\frac{\theta_{II}}{2}\right)^2,
\end{eqnarray}
\begin{eqnarray}\label{cII}
 c_{II}&=&\gamma_{II,11}\left(\cos\frac{\theta_I}{2}\sin\frac{\theta_{II}}{2}
 -\sin\frac{\theta_I}{2}\cos\frac{\theta_{II}}{2}\right)^2\nonumber\\
 \nonumber\\
 &+&\gamma_{II,22}\left(\cos\frac{\theta_I}{2}\cos\frac{\theta_{II}}{2}
 -\sin\frac{\theta_I}{2}\sin\frac{\theta_{II}}{2}\right)^2,
\end{eqnarray}
\begin{eqnarray}\label{cIcross}
 c_{cr,I}&=&\gamma_{I,11}\left(\cos\frac{\theta_I}{2}\cos\frac{\theta_{II}}{2}
 +\sin\frac{\theta_I}{2}\sin\frac{\theta_{II}}{2}\right)^2\nonumber\\
 \nonumber\\
 &-&\gamma_{I,22}\left(\cos\frac{\theta_I}{2}\sin\frac{\theta_{II}}{2}
 +\sin\frac{\theta_I}{2}\cos\frac{\theta_{II}}{2}\right)^2,
\end{eqnarray}
\begin{eqnarray}\label{cIIcross}
 c_{cr,II}&=&-\gamma_{II,11}\left(\cos\frac{\theta_I}{2}\sin\frac{\theta_{II}}{2}
 -\sin\frac{\theta_I}{2}\cos\frac{\theta_{II}}{2}\right)^2\nonumber\\
 \nonumber\\
 &+&\gamma_{II,22}\left(\cos\frac{\theta_I}{2}\cos\frac{\theta_{II}}{2}
 -\sin\frac{\theta_I}{2}\sin\frac{\theta_{II}}{2}\right)^2,
\end{eqnarray}

The corresponding excitation rates $\bar{c}_i$, and the cross terms
$\bar{c}_{cr,i}$, are obtained by substituting, in Eqs.
(\ref{cI})-(\ref{cIIcross}), $\gamma_{i,ll}$ with the corresponding
quantities $\bar{\gamma}_{i,ll}$: when the temperatures of the two
reservoirs $T_1$ and $T_2$ are both zero, all this coefficients
vanish, according to Eq. (\ref{KMS}). Physically this means that
there is no possibility to create excitations in the bipartite
system due to the interaction with the reservoirs.

Let us conclude this section by reminding that the master equation
has been derived in the framework of Born-Markov and rotating wave
approximations. The last approximation is valid as long as the
relaxation time of the system is much longer than the time
characterizing its unitary dynamics \cite{petruccionebook}.
Mathematically this is equivalent to assume that the coupling with
the environment is weak enough to say that the relaxation rates
are all much smaller than the smallest nonzero Bohr frequency
relative to $H_S$, that is $\omega_I$.


\section{Dynamics}

The master equation given by Eq. (\ref{Meqfinale}), is equivalent
to a system of coupled differential equation, the first four of
which describe the time evolution of the populations of the
dressed states $\ket{a}$, $\ket{b}$, $\ket{c}$ and $\ket{d}$,
namely

\begin{eqnarray}\label{EQpopulation}
\dot{\rho}_{aa}(t)=-\left(\bar{c}_I+\bar{c}_{II}\right)\rho_{aa}(t)+c_I\rho_{bb}(t)+c_{II}\rho_{cc}(t) \nonumber\\
 \nonumber\\
\dot{\rho}_{bb}(t)=\bar{c}_I\rho_{aa}(t)-\left(c_I+\bar{c}_{II}\right)\rho_{bb}(t)+c_{II}\rho_{dd}(t) \nonumber\\
 \nonumber\\
\dot{\rho}_{cc}(t)=\bar{c}_{II}\rho_{aa}(t)-\left(c_{II}+\bar{c}_{I}\right)\rho_{cc}(t)+c_I\rho_{dd}(t)\nonumber\\
 \nonumber\\
\dot{\rho}_{dd}(t)=\bar{c}_{II}\rho_{bb}(t)+\bar{c}_I\rho_{cc}(t)-\left(c_I+c_{II}\right)\rho_{dd}(t),
\end{eqnarray}
while the other equations describe the time evolution of the
coherences:
\begin{eqnarray}\label{EQcoherenceII}
\dot{\rho}_{ac}(t)=\left[i\omega_{II}
-\frac{\left(c_{II}+2\bar{c}_I+\bar{c}_{II}\right)}{2}\right]\rho_{ac}(t)+c_{cr,I}\rho_{bd}(t)\nonumber\\
 \nonumber\\
\dot{\rho}_{bd}(t)=\left[i\omega_{II}
-\frac{\left(2c_I+c_{II}+\bar{c}_{II}\right)}{2}\right]\rho_{bd}(t)+\bar{c}_{cr,I}\rho_{ac}(t)
\end{eqnarray}

\begin{eqnarray}\label{EQcoherenceI}
\dot{\rho}_{ab}(t)=\left[i\omega_I
-\frac{\left(c_I+\bar{c}_I+2\bar{c}_{II}\right)}{2}\right]\rho_{ab}(t)+c_{cr,II}\rho_{cd}(t)\nonumber\\
 \nonumber\\
\dot{\rho}_{cd}(t)=\left[i\omega_I
-\frac{\left(c_I+2c_{II}+\bar{c}_I\right)}{2}\right]\rho_{cd}(t)+\bar{c}_{cr,II}\rho_{ab}(t)
\end{eqnarray}

\begin{eqnarray}\label{EQcoherencefree}
\dot{\rho}_{ad}(t)=\left[i\omega_{da}
-\frac{\left(c_I+c_{II}+\bar{c}_I+\bar{c}_{II}\right)}{2}\right]\rho_{ad}(t)\nonumber\\
 \nonumber\\
\dot{\rho}_{bc}(t)=\left[i\omega_{cb}
-\frac{\left(c_I+c_{II}+\bar{c}_I++\bar{c}_{II}\right)}{2}\right]\rho_{bc}(t),
\end{eqnarray}
where $\omega_{da}=E_d-E_a$ and $\omega_{cb}=E_c-E_b$. The other
coherences can be obtained by complex conjugation. In the
following subsections we will discuss the existence of a
stationary solution and we will bring to light the features
characterizing the dynamics of the entanglement when both $T_1$
and $T_2$ are equal to zero.

\subsection{Solution I: The stationary state}

Let us look for the existence of the stationary solution by
imposing $\dot{\rho}_{ii}=0$ ($\forall \,i\equiv \{a,b,c,d\}$) in
Eqs. (\ref{EQpopulation}) and the normalization condition
$Tr(\rho(t))=1$.  In such a condition, after some calculations, it
is possible to prove the existence of a stationary solution given
by:
\begin{eqnarray}\label{soluzionestazionaria}
\rho_{aa,\,ST}=\frac{c_Ic_{II}}{(c_I+\bar{c}_I)(c_{II}+\bar{c}_{II})} \nonumber\\
\rho_{bb,\,ST}=\frac{\bar{c}_Ic_{II}}{(c_I+\bar{c}_I)(c_{II}+\bar{c}_{II})} \nonumber\\
\rho_{cc,\,ST}=\frac{c_I\bar{c}_{II}}{(c_I+\bar{c}_I)(c_{II}+\bar{c}_{II})} \nonumber\\
\rho_{dd,\,ST}=\frac{\bar{c}_I\bar{c}_{II}}{(c_I+\bar{c}_I)(c_{II}+\bar{c}_{II})}
\end{eqnarray}
As it is immediate to see, when $T_1=T_2=0$ K, $\rho_{aa,\,ST}=1$,
namely the stationary solution coincides with the ground state of
the bipartite system. In view of Eq. (\ref{eigenstates}), this
implies the existence of a stationary entanglement traceable back
to the presence of counter-rotating terms in the interaction
Hamiltonian (see Eq. (\ref{Hamiltonianmodel})) describing the
coupling between the two two-state systems. We will discuss
stationary entanglement in the next subsection.

\subsection{Solution II: Entanglement dynamics at zero temperature}

Let us consider now the analysis of the system dynamics when the
temperatures of the two reservoirs, $T_1$ and $T_2$, are equal to
$0$ K. Moreover let us restrict our attention to the case of exact
resonance $\omega_1=\omega_2$ and in which the two environment
have both flat spectra and are coupled with equal strength to the
respective subsystems, which means that
$\gamma_{I,11}=\gamma_{II,11}=\gamma_{I,22}=\gamma_{I,22}$. From
Eqs. (\ref{thetaII}), (\ref{cIcross}) and (\ref{cIIcross}), it is
straightforward to show that in this case the cross coefficients
$c_{cr,I}$ and $c_{cr,II}$ are both equal to zero, so that each
coherence between the dressed states of the system evolves
independently for the other ones.

Putting this condition into Eqs. (\ref{EQpopulation})
-(\ref{EQcoherencefree}), it is possible to show that their
solution is:
\begin{eqnarray}\label{roaat}
\rho_{aa}(t)=1-\rho_{bb}(t)-\rho_{cc}(t)-\rho_{dd}(t),
\end{eqnarray}
\begin{eqnarray}\label{robbt}
\rho_{bb}(t)=\left(\rho_{bb}(0)+\rho_{dd}(0)\right)\mathrm{e}^{-c_{I}t}-\rho_{dd}(0)\mathrm{e}^{-(c_I+c_{II})t},
\end{eqnarray}
\begin{eqnarray}\label{rocct}
\rho_{cc}(t)=\left(\rho_{cc}(0)+\rho_{dd}(0)\right)\mathrm{e}^{-c_{II}t}-\rho_{dd}(0)\mathrm{e}^{-(c_I+c_{II})t},
\end{eqnarray}
\begin{eqnarray}\label{roddt}
\rho_{dd}(t)=\rho_{dd}(0)\mathrm{e}^{-(c_I+c_{II})t}
\end{eqnarray}
 and
\begin{eqnarray}\label{coheAC}
\rho_{ac}(t)=\mathrm{e}^{\left[i\omega_{II}
-\frac{\left(c_{II}+2\bar{c}_I+\bar{c}_{II}\right)}{2}\right]t}\rho_{ac}(0)
\end{eqnarray}
\begin{eqnarray}\label{coheBD}
\rho_{bd}(t)=\mathrm{e}^{\left[i\omega_{II}
-\frac{\left(2c_I+c_{II}+\bar{c}_{II}\right)}{2}\right]t}\rho_{bd}(0)
\end{eqnarray}
\begin{eqnarray}\label{coheAB}
\rho_{ab}(t)=\mathrm{e}^{\left[i
\omega_I-\frac{\left(c_I+\bar{c}_I+2\bar{c}_{II}\right)}{2}\right]t}\rho_{ab}(0)
\end{eqnarray}
\begin{eqnarray}\label{coheCD}
\rho_{cd}(t)=\mathrm{e}^{\left[i\omega_I
-\frac{\left(c_I+2c_{II}+\bar{c}_I\right)}{2}\right]t}\rho_{cd}(0)
\end{eqnarray}
\begin{eqnarray}\label{coheAD}
\rho_{ad}(t)=\mathrm{e}^{\left[i\omega_{da}
-\frac{\left(c_I+c_{II}+\bar{c}_I+\bar{c}_{II}\right)}{2}\right]t}\rho_{ad}(0)
\end{eqnarray}
\begin{eqnarray}\label{coheBC}
\rho_{bc}(t)=\mathrm{e}^{\left[i\omega_{cb}
-\frac{\left(c_I+c_{II}+\bar{c}_I++\bar{c}_{II}\right)}{2}\right]t}\rho_{bc}(0)
\end{eqnarray}
\\
Starting from the initial state
\begin{eqnarray}\label{stato_iniziale_1ecc_gen}
\ket{\psi(0)}=\sqrt{p}\ket{01}+\mathrm{e}^{i
\varphi}\sqrt{1-p}\ket{10},
\end{eqnarray}
$p$ being a non-negative real number $\leq 1$, and by taking into
account that
\begin{eqnarray}\label{rhobbzero1}
\rho_{bb}(0)=\left\vert\mathrm{e}^{i \varphi}\sqrt{1-p}\,
\cos\left(\frac{\theta_{II}}{2}\right)-\sqrt{p}\,\sin\left(\frac{\theta_{II}}{2}\right)\right\vert^2,
\end{eqnarray}
\begin{eqnarray}\label{rhocczero1}
\rho_{cc}(0)=\left\vert\mathrm{e}^{i \varphi}\sqrt{1-p}\,
\sin\left(\frac{\theta_{II}}{2}\right)+\sqrt{p}\,\cos\left(\frac{\theta_{II}}{2}\right)\right\vert^2,
\end{eqnarray}
\begin{eqnarray}\label{rhobczero1}\nonumber
\rho_{bc}(0)&=&\left(\mathrm{e}^{i \varphi}\sqrt{1-p}\,
\cos\left(\frac{\theta_{II}}{2}\right)-\sqrt{p}\,\sin\left(\frac{\theta_{II}}{2}\right)\right)\\
&&\times\left(\mathrm{e}^{-i \varphi}\sqrt{1-p}\,
\sin\left(\frac{\theta_{II}}{2}\right)+\sqrt{p}\,\cos\left(\frac{\theta_{II}}{2}\right)\right),
\end{eqnarray}
while $\rho_{ii}=0$ and $\rho_{ij}=0$ otherwise, it is possible to
see that the density matrix describing the time evolution of the
bipartite system at a generic instant of time $t$ assumes the
following form:
\begin{eqnarray}\label{evoluz1}\nonumber
\rho(t)&=&\rho_{aa}(t)\ket{a}\bra{a}+\rho_{bb}(t)\ket{b}\bra{b}+\rho_{cc}(t)\ket{c}\bra{c}
+\rho_{dd}(t)\ket{d}\bra{d}\\
&&+\rho_{bc}(t)\ket{b}\bra{c}+\rho_{cb}(t)\ket{c}\bra{b}
\end{eqnarray}
To assess how much entanglement is stored in this bipartite
quantum system at different instants of time we use the
concurrence, a function introduced by Wootters \cite{Wotters},
equal to $1$ for maximally entangled states and zero for separable
states, defined as:
\begin{eqnarray}\label{concurrence}
\mathcal{C}(t)=max(0,\sqrt{\lambda_1(t)}-\sqrt{\lambda_2(t)}-\sqrt{\lambda_3(t)}-\sqrt{\lambda_4(t)})
\end{eqnarray}
where $\{\lambda_i(t)\}$ are the eigenvalues of the matrix
\begin{eqnarray}\label{R}
R(t)=\rho(t) \tilde{\rho}(t),
\end{eqnarray}
 with $\tilde{\rho}(t)$ given by
$\tilde{\rho}(t)=\sigma_y\otimes\sigma_y\rho^*(t)\sigma_y\otimes\sigma_y$,
$\sigma_y$ being the Pauli matrix, and $\rho(t)$ the density
matrix representing the quantum state of the system. Inserting Eq.
(\ref{evoluz1}) into Eq. (\ref{R}), it is possible to derive the
time evolution of concurrence, studying in this way the features
of the non-classical correlations characterizing the system of the
two coupled qubits. More in details, it is possible to see that
when the system is initially prepared in the factorized state
$\ket{01}$ its dynamics is characterized by the existence of
damped oscillations corresponding to the periodic appearance and
disappearance of non-classical correlations between the two
qubits. The oscillations in the concurrence are a direct signature
of the well known oscillations of the single excitation between
the two qubits, which are caused by the resonant terms in the
qubit-qubit interaction. On the other hand the stationary
entanglement at long times is a consequence of the fact that, due
to the counter-rotating terms in the Hamiltonian in Eq.
(\ref{Hamiltonianmodel}), the ground state of the system is not
the unexcited state $\left|00\right>$, but the entangled state
$\ket{a}$ in Eq. (\ref{eigenstates}). From Eq. (\ref{thetaI}) it
is straightforward to see that the smaller the coupling constant
$\lambda$ the smaller the amount of stationary entanglement.

The system dynamics is even richer when the system is initially
prepared in a state with two excitations or in an arbitrary
superposition of the states $\ket{00}$ and $\ket{11}$, i.e.
\begin{eqnarray}\label{stato_iniziale_2ecc_gen}
\ket{\psi(0)}=\sqrt{p}\ket{00}+\mathrm{e}^{i
\varphi}\sqrt{1-p}\ket{11}.
\end{eqnarray}
Also in this case $p$ is a non-negative real number $\leq 1$,
while we have:
\begin{eqnarray}\label{rhoaazero2}
\rho_{aa}(0)=\left\vert\sqrt{p}\,\cos\left(\frac{\theta_{I}}{2}\right)-\mathrm{e}^{i
\varphi}\sqrt{1-p}\,
\sin\left(\frac{\theta_{I}}{2}\right)\right\vert^2,
\end{eqnarray}
\begin{eqnarray}\label{rhoddzero2}
\rho_{dd}(0)=\left\vert\sqrt{p}\,\sin\left(\frac{\theta_{I}}{2}\right)+\mathrm{e}^{i
\varphi}\sqrt{1-p}\,
\cos\left(\frac{\theta_{I}}{2}\right)\right\vert^2,
\end{eqnarray}
\begin{eqnarray}\label{rhoadzero2}\nonumber
\rho_{ad}(0)&=&\left(\sqrt{p}\,\cos\left(\frac{\theta_{I}}{2}\right)-\mathrm{e}^{i
\varphi}\sqrt{1-p}\,
\sin\left(\frac{\theta_{I}}{2}\right)-\right)\\
&&\times\left(\sqrt{p}\,\sin\left(\frac{\theta_{I}}{2}\right)+\mathrm{e}^{-i
\varphi}\sqrt{1-p}\, \cos\left(\frac{\theta_{I}}{2}\right)\right),
\end{eqnarray}
with $\rho_{ii}=0$ and $\rho_{ij}=0$ otherwise. With such an
initial condition, the density matrix of the system at a generic
instant of time assumes the form
\begin{eqnarray}\label{evoluz2}\nonumber
\rho(t)&=&\rho_{aa}(t)\ket{a}\bra{a}+\rho_{bb}(t)\ket{b}\bra{b}+\rho_{cc}(t)\ket{c}\bra{c}
+\rho_{dd}(t)\ket{d}\bra{d}\\
&&+\rho_{ad}(t)\ket{b}\bra{c}+\rho_{da}(t)\ket{c}\bra{b}.
\end{eqnarray}
Exploiting Eq. (\ref{stato_iniziale_2ecc_gen}) (with $p=0$) as well
as Eqs. (\ref{rhoaazero2})-(\ref{evoluz2}) it easy to derive the
time evolution of the concurrence of the two qubits. Figure 3 shows
that, apart from the oscillations of the entanglement due to the
oscillations of the excitation in the one-excitation subspace, the
behaviour of the concurrence is characterized by the phenomena of
sudden birth and sudden death of the entanglement already found by
Ficek and Tanas \cite{Tanas} for the scenario of two interacting
qubits. While the phenomenon entanglement sudden death, i.e., the
complete disentanglement of the system in a finite time, is quite
well understood and is a feature common to any dissipative
two-qubits dynamics, provided the system starts from the proper
initial state \cite{Yu,bellomo}, the occurrence of stationary
entanglement is a phenomenon usually ascribed to the presence of a
reservoir which is common to the two parts of the bipartite system
\cite{Tanas,manu}. What we have proved here is that the same
phenomenon can occur even in the presence of independent reservoirs
for the two qubits, provided the interaction Hamiltonian between the
two qubits contains also the counter-rotating terms which are
usually neglected in the study of the dynamics. This is due to the
fact that, in the weak damping limit, the quantum jumps occur
between eigenstates of the system Hamiltonian and to the fact that,
due to the presence of counter-rotating terms in the two-qubit
interaction, the ground state of the system is an entangled state,
as we have seen before.


\begin{figure}[hb!]
 \begin{center}
     \includegraphics[width=0.7\textwidth,
height=0.40\textwidth]{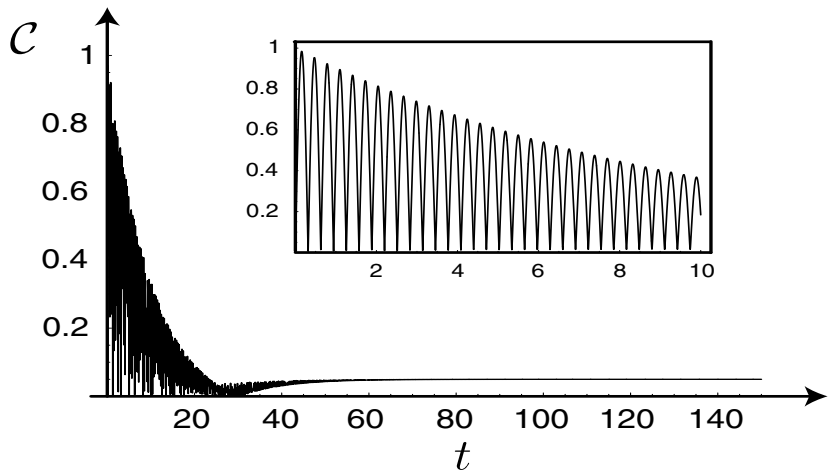}
    \caption{Concurrence of the two qubits as a function of time $t$ (in units of $10*(\lambda)^{-1})$.
    The initial state is $\ket{01}$, the parameters assume the values $\omega_1/\lambda=\omega_2/\lambda=10$,
     $T_1=T_2=0$ K, $\gamma_{I,11}/\lambda=\gamma_{II,11}/\lambda=\gamma_{I,22}/\lambda=\gamma_{I,22}/\lambda=0.01$.}\label{figura1}
 \end{center}
\end{figure}



\begin{figure}[hb!]
 \begin{center}
     \includegraphics[width=0.7\textwidth,
height=0.40\textwidth]{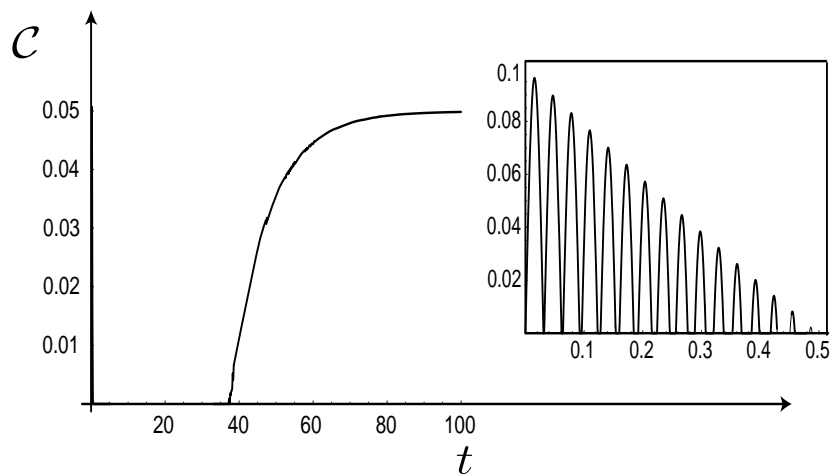}
    \caption{Concurrence of the two qubits as a function of time $t$ (in units of $10*(\lambda)^{-1})$.
    The initial state is $\ket{11}$, the parameters assume the values $\omega_1/\lambda=\omega_2/\lambda=10$,
     $T_1=T_2=0$ K, $\gamma_{I,11}/\lambda=\gamma_{II,11}/\lambda=\gamma_{I,22}/\lambda=\gamma_{I,22}/\lambda=0.01$.}\label{figura2}
 \end{center}
\end{figure}


\section{Discussion and Conclusive Remarks}

To summarize, we have presented a microscopic derivation of the
Markovian master equation in the weak damping limit governing the
time evolution of two interacting two-state systems, each of them
coupled to independent bosonic reservoirs, and we have studied the
time evolution of the entanglement between the two qubits for
various initial conditions of the bipartite system.

The behaviour of the concurrence is consistent with the results of
the literature, showing the occurrence of entanglement sudden
death for some initial states and of stationary entanglement for
any initial state. The presence of stationary entanglement which
is not washed away from dissipation is perhaps the most important
result of our analysis. Indeed this feature is usually ascribed to
the presence of a common reservoir which correlates the two parts
of the bipartite system. Instead we have shown here that
stationary entanglement can occur also due to the interaction
between the two parts, provided their interaction is described in
a complete way which includes the counter-rotating terms in their
interaction Hamiltonian. We feel that our approach could help in
understanding the role of energy non-conserving terms in the
dissipative dynamics of open bipartite quantum systems. From this
point of view more is expected from the analysis of the cases with
reservoirs at different temperatures or with non-flat spectrum, in
the framework of a non-Markovian extension of the master equation
presented here. These points will be the subject of our future
work.

\section*{Acknowledgements}
M.S. thanks the Fondazione Angelo Della Riccia for financial
support. A.M. acknowledges partial support by MIUR project
II04C0E3F3 \textit{Collaborazioni Interuniversitarie ed
Internazionali Tipologia C}.

\end{document}